\documentclass[10pt,final]{article}
\usepackage{graphicx}
\usepackage{amssymb}
\usepackage{amsmath}
\usepackage{bm}
\usepackage{footmisc}

\usepackage[superscript,biblabel]{cite}
\usepackage{multicol,caption}
\newenvironment{Figure}
{\par\medskip\noindent\minipage{\linewidth}}
{\endminipage\par\medskip}

\begin{document}

\vspace*{1cm}

{\sffamily\huge\bfseries\noindent Giant and accessible conductivity of charged domain
walls in lithium niobate}

\vspace{0.5cm} \renewcommand{\thefootnote}{\fnsymbol{footnote}}
{\sffamily\small\noindent Christoph S. Werner$^{1,}$\footnote[1]{These authors contributed equally to this work.},\renewcommand*{\thefootnote}{\arabic{footnote}} Simon J.\ Herr\footnote{Department of Microsystems Engineering - IMTEK, University of Freiburg, Georges-K\"{o}hler-Allee 102, 79110 Freiburg, Germany.}$^{,*}$, Karsten Buse$^{1,}$\footnote{Fraunhofer Institute for Physical Measurement Techniques IPM, Heidenhofstra\ss e 8, 79110 Freiburg, Germany.}, Boris Sturman\footnote{Institute for Automation and Electrometry of Russian Academy of Science, 630090 Novosibirsk, Russia.},
	Elisabeth Soergel\footnote{Institute of Physics, University of Bonn, Wegelerstra\ss e 8, 53115 Bonn, Germany.}, Cina Razzaghi$^4$, and Ingo Breunig$^{1,2}$}

\vspace*{0.3cm}

\begin{multicols}{2}

\noindent {\fontseries{b}\selectfont \small Ferroelectric domain walls are nm-sized
interfaces between sections of different allowed values of the spontaneous
polarization, the so-called domains. These walls -- neutral or charged -- can be created, displaced, deleted,
and recreated again in ferroelectric materials\cite{CatalanRMP12,BednyakovSciRep15,FeiglSciRep16}.
Owing to the recent progress in the studies of ferroelectrics, they are expected to be functional active elements of the future
nano-electronics\cite{CatalanRMP12,SeidelPCL12,VasudevanAFM13}. Metallic-like
conductivity of charged domain walls (CDWs) in insulating ferroelectrics, predicted in
$\bm{1970}$s\cite{Vul73} and detected
recently\cite{Maksymovych12,VasudevanNL12,SlukaNC13,Chapter16}, is especially attractive
for applications. This important effect is still in its infancy. The electric currents
are small, the access to the conductivity is hampered by contact barriers, and stability
is low because of sophisticated domain structures and/or proximity of the Curie point.
Here, we report on giant and accessible persistent CDW conductivity in lithium niobate (LN)
crystals (LiNbO$_3$) -- a vital material for photonics\cite{Arizmendi04}. Our results
are by far superior to the data known for other materials: Increase of LN conductivity
by more than 14 orders of magnitude owing to CDWs, access to the effect via Ohmic and
semi-Ohmic contacts, and its high stability for temperatures of up to
$\bm{70^{\,\circ}}$C are demonstrated. It is made clear why this big effect was missed
earlier\cite{SoergelAFM12,ShurAPL13} in LN. Our results demonstrate that strong
conductivity of CDWs is available in simple ferroelectric materials, possessing only two
allowed orientations of the spontaneous polarization, far from the Curie point. Also,
CDW functionalities can be combined with linear and nonlinear optical
phenomena\cite{Dunn99,WeigelSciRep16}. Our findings allow new generations of
adaptive-optical elements, of electrically controlled integrated-optical chips for
quantum photonics, and of advanced LN-semiconductor hybrid optoelectronic devices. }


\noindent {\fontseries{c}\selectfont \small The ferroelectric state can be characterized
by the spatial distribution of the spontaneous polarization vector $\bm{P}_s$. Within a
ferroelectric domain $\bm{P}_s \simeq {\rm const}$. Domain walls separate the regions
with different allowed polarizations. Often, they are neutral, then the polarization
difference $\Delta \bm{P}_s$ is parallel to the wall plane and the bound polarization
charge is zero. Otherwise, the walls are charged and the bound charge is strongly
compensated by free charges to ensure the ferroelectric
stability\cite{Vul73,GurievPRB11,Zuo14}. A simple example of CDW, relevant to our study,
is depicted in Fig.~\ref{fig1}a. The polarization vectors of the adjacent domains point at
the wall (head-to-head CDW), which is tilted by an angle $\theta$ to the polarization
direction. Here, the bound surface charge is $+2P_s\sin \theta$, and it must be strongly
compensated by electrons or negative ions. The electron compensation occurs via the
conduction band bending~\cite{Vul73,Chapter16,GurievPRB11}. It is fast and leads to
a metallic-like CDW conductivity because of free movement of electrons along the wall.
The width $w$ of CDWs can roughly be estimated as $10$~nm, this is one order of
magnitude larger than the width of neutral walls\cite{Chapter16,GurievPRB11}. For LN
polarization $P_s = 70\,\mu$C/cm$^2$ and $\theta = 90^{\circ}$ it gives a huge
concentration of electrons, $2P_s/ew \approx 10^{21}$~cm$^{-3}$, where $e$ is the
elementary charge.

\begin{Figure}
\centering
\includegraphics[width=8.3cm,height=7cm]{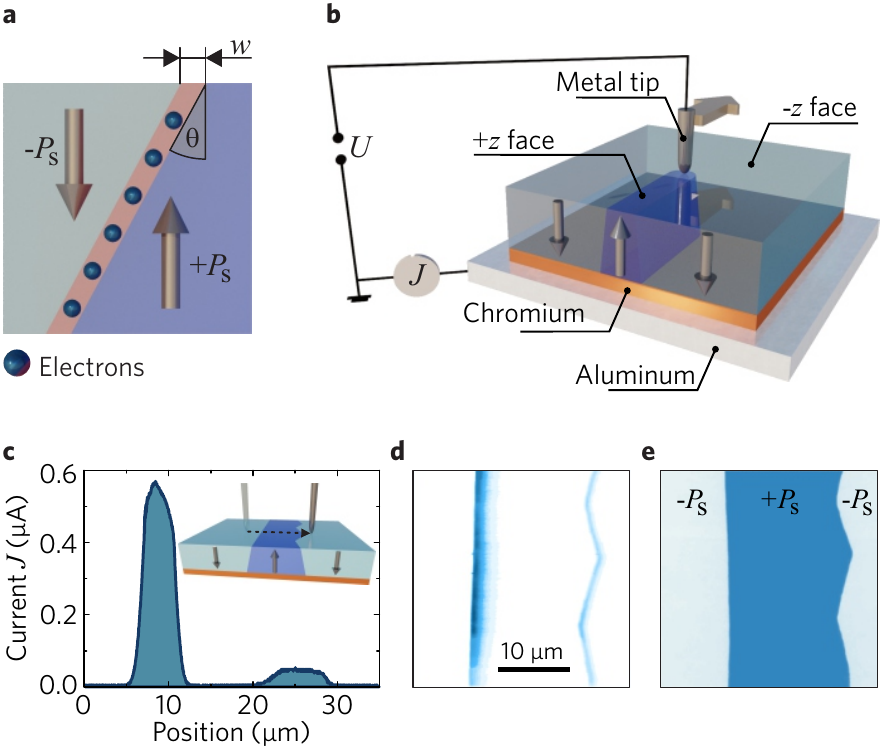}
\captionof{figure}{{\bf Head-to-head charged domain walls in LN crystals and tip-related experiments on
CDW conduction.} {\bf a,} Schematic of a head-to-head CDW in LN crystals; only two
opposite directions of $\bm{P}_s$ are allowed. {\bf b,} Schematic of experimental setup
for recording and investigation of CDWs in LN crystals. The vertical arrows show
directions of $\bm{P}_s$. The inverted domain broadens from top to bottom. {\bf c,}
The current $J$ versus the tip position about the domain line during a readout
experiment at $U = 50$~V. {\bf d,} The spatial distribution $J(x,y)$ corresponding to
Fig.~\ref{fig1}c obtained with an AFM tip. {\bf e,} The corresponding PFM image
identifying the $\pm z$ domains.}\label{fig1}
\end{Figure}

CDWs and their elevated conductivity were occasionally observed in the
past~\cite{Grekov76,SeidelNM09}. During the last decade, they were well documented in
many materials including BiFeO$_3$\cite{Crassous15},
BaTiO$_3$\cite{SlukaNC13,Bednyakov16}, ErMnO$_3$\cite{MeierNM12}, and
h-HoMnO$_3$\cite{Wu12}. High-resolution transmission-electron and atomic-force
microscopies were extensively used. The most advanced data on the CDW conductivity has been
presented so far for BaTiO$_3$ single crystals. An enhancement factor $\sim 10^9$, as
compared to the conductivity of the bulk, was obtained for $45^{\circ}$ head-to-head
walls. The data reported suffer from bad stability and big contact barriers leading to
pronounced non-Ohmic current-voltage characteristics. Experiments with LN crystals have
shown either transient conductivity\cite{ShurAPL13} or steady conductivity just in the
presence of super-band-gap illumination\cite{SoergelAFM12}.

Our choice of LN crystals for domain-wall-electronic studies has strong grounds. This
robust and cheap wide-band-gap material ($E_g \simeq 4$~eV) is extremely useful for
numerous optical applications\cite{Arizmendi04,Dunn99}. As a ferroelectric, it has a very
high Curie temperature, $T_C \approx 1200^{\,\circ}$C, and only two allowed (opposite)
values of $\bm{P}_s$\cite{LNProperties}. Field-assisted domain engineering is a
well-developed area because of the quasi-phase-matching
applications\cite{Arizmendi04,LNPoling,CallPoling05}. The dark conductivity of LN
crystals is extremely low\cite{LNProperties}.

In our experiments, we used mainly $300$-$\mu$m-thick $z$-cut samples of congruent LN
doped with $5$ mol.\%~MgO. The data presented in this article correspond to this material.
However, similar data were obtained with near-stoichiometric LN. In order to produce
domain structures in originally single-domain crystals, we employed the calligraphic domain writing\cite{CallPoling05}. It is shown schematically in
Fig.~\ref{fig1}b. The bottom $+z$ face was coated with a $0.5$-$\mu$m-thick Cr electrode
and then glued with conductive paste to a heatable aluminum mount. The temperature $T$ can be
controlled on the $0.1^{\,\circ}$C level within the range $(20 - 150)\,^{\circ}$C. As
the top electrode we use a tungsten-carbide needle with $\mu$m-sized tip radius. During
the domain-inversion procedure, the tip is moving slowly along the $x$
direction producing a domain line which is typically as long as $1$~mm. The applied
voltage $U$ is kept such that the poling current $J \simeq 30$~nA. Subsequent etching of
the sample shows that the domain size on the $+z$ face of the untreated crystal noticeably and reproducibly
exceeds the size on its $-z$ face. Thus, the domain walls are slightly tilted, and they
must be positively charged. The tilt angle can be
estimated as $\theta \approx 1^{\circ}$. Remarkably, the total charge flowed through the
sample during the domain-inversion procedure exceeds the net polarization charge by
orders of magnitude. This is an indication of CDW conductivity. The major experimental differences compared to previous work are the use of room temperature poling and the avoidance of after-poling heat
treatment.

In the first series of experiments on the prepared domain structure, a virgin
$\mu$m-sized tip is moving slowly across the domain line, as shown in the inset of
Fig.~\ref{fig1}c. The voltage is decreased to $50$~V to reduce the applied field far
below the coercive field ($E_c \gtrsim 3$~kV/mm) and hence to avoid any field effect on the
domain structure. The current is nonzero only when crossing the left and right domain
borders; this proves that CDWs are responsible for the conduction. The value $J_{\max}
\simeq 0.55\,\mu$A is very high for tip-related experiments. The relative asymmetry
between the left and right borders in Fig.~\ref{fig1}c is occasional -- it varies
 when crossing different sections of the domain line. Employment
of a conductive AFM tip of about $50$~nm radius has allowed us to map the current
distribution in the same area, see Fig.~\ref{fig1}d. Within the actual spatial
resolution, the conductivity is strongly localized at the left CDW. Relative smallness
of the current near the right wall correlates with its zigzag structure. The PFM image
of the same area, see Fig.~\ref{fig1}e, reproduces the straight left border of the
domain and its zigzag right border.

Next, we kept our $\mu$m-sized tip at the point of maximum of $J(y)$ and measured the
current-voltage characteristics. The corresponding result is presented in
Fig.~\ref{fig2}.
\begin{Figure}
\centering
\includegraphics[width=8.6cm]{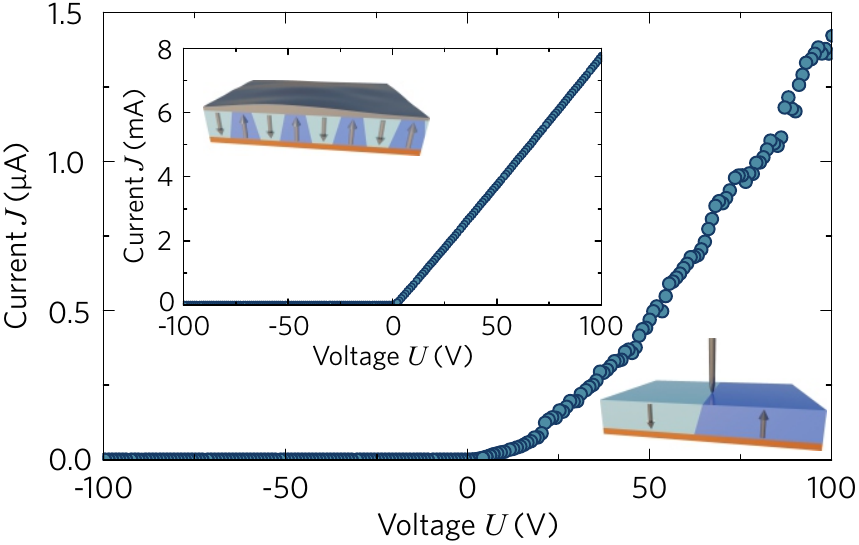}
\captionof{figure}{Diode-like current-voltage dependence at the point of maximum of
$J(y)$ for a single domain line. The inset corresponds to an array of $732$ lines covered
with a silver-paste top electrode. The tilts of the walls are exaggerated. }\label{fig2}
\end{Figure}
\noindent A pronounced diode-like dependence of $J(U)$ is clearly seen. For $U < 0$ the
current is practically absent, while for $U \gtrsim 30$~V we have an almost quasi-Ohmic behavior. While the Schottky-like $J(U)$ characteristic is typical for a tip
experiment, the quasi-Ohmic behavior is rather special. The above $J(U)$
measurement has then been advanced. Instead of a single domain line we have recorded an
array of $732$ parallel $1$-mm-long domain lines. After that, the top face was covered
with conductive silver paste. Application of voltage leads here to a strikingly regular
semi-Ohmic $J(U)$ characteristic, see the inset in Fig.~\ref{fig2}. For $U > 0$, this
characteristic is practically linear showing the absence of electrode barriers. The
values of the current approach here $0.01$~A.

The above data allow us to evaluate reliably the CDW conductivity. To make so, we
calculate the total resistance of the system $R = U/J$ under the assumption of Ohmic
contacts. Since the problem is two-dimensional, we have generally for a single wall:
$R_1 = \rho_s F$, where $\rho_s = (\sigma w)^{-1}$, $\rho_s$ is the specific surface CDW
resistivity, $\sigma$ is the average conductivity inside the wall, and $F$ is a
dimensionless factor determined by the geometry of the contacts\cite{Smythe}. Let $L$ and
$a$ be the domain-line length and the electrode length along the wall, respectively. For
a continuous top electrode we have $a \simeq L \gg d$. Here, the stream lines of the current
are parallel to each other and, by analogy with the plane capacitor, we have $F \simeq
d/L$. For $N$ parallel-connected CDWs we have $R \simeq \rho_s d/NL$. Setting $d =
300\,\mu$m, $L = 1$~mm, $N = 732$, and estimating $V/J$ from the inset of
Fig.~\ref{fig2}, we obtain $\rho_s \simeq 3.1 \times 10^7\,\Omega$. In the tip-electrode
case, the stream lines of the current are expanding in the wall plane with the distance from
the top electrode. Numerical simulations give here $F \simeq 0.72\, {\rm lg}(5d/a)$.
This logarithmic dependence on $d/a$ is fairly weak. Setting $a = 2\,\mu$m, which is not
far from the tip diameter, and estimating $R_1 = U/J$ from Fig.~\ref{fig2}, we obtain
$\rho_s \approx 3.5 \times 10^7\,\Omega$. The consistency of the two independent estimates,
obtained for different contacts and geometries, is striking. Setting next $w = 10$~nm,
we obtain for the bulk CDW conductivity: $\sigma \approx 0.03$~$(\Omega \, {\rm
cm})^{-1}$. This value exceeds the dark conductivity of LN crystals\cite{LNProperties}
by at least $14$ orders of magnitude. Finally, using the relation $\sigma = e\mu_e n$
and setting for the electron drift mobility $\mu_e \lesssim 0.1$~cm$^2$/Vs, we obtain for the concentration of
compensating electrons: $n \gtrsim 1.8 \times 10^{18}$~cm$^{-3}$. This is in fair
agreement with the estimate $n \approx 2P_s\theta/e w$ for $\theta \approx 1^{\circ}$.

Our experimental efforts are directed next to elucidate the origin of the contact
barriers and to get access to the CDW conductivity through fully Ohmic contacts. To
accomplish this challenging task, we have recorded an array of sixteen 13\hbox{-}mm\hbox{-}long
domain lines. Then the central part of the bottom Cr electrode has been etched away, and
two spatially separated droplets of conductive paste are used as two top electrodes, see
Fig.~\ref{fig3}a.
\begin{Figure}
\centering
\includegraphics[width=8.6cm]{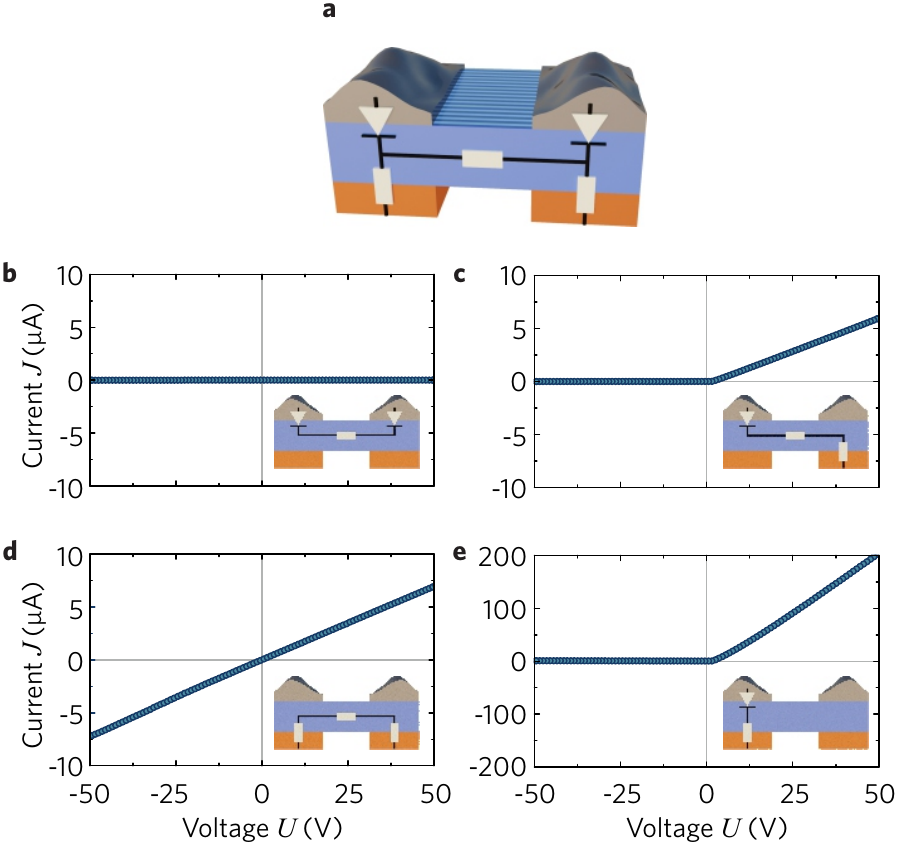}
\captionof{figure}{{\bf Origin of contact barriers and Ohmic access to CDW
conductivity.} {\bf a,} Geometry of four-electrode configuration and an equivalent
electric scheme. {\bf b,} Top-Top connection: zero current. {\bf c,}
Top-left-Bottom-right connection: semi-Ohmic behavior. {\bf d,} Bottom-Bottom
connection: fully Ohmic behavior. {\bf e,} Top-left-Bottom-left connection: semi-Ohmic
behavior. }\label{fig3}
\end{Figure}
\noindent Zero current was observed when the voltage was applied between two top
electrodes, see Fig.~\ref{fig3}b. This shows unambiguously that the top electrodes act
as almost perfect top-bottom diodes. Applying voltage between the left-top and
right-bottom electrodes or between right-top and left-bottom electrodes
(Figs.~\ref{fig3}c and~\ref{fig3}e) returns us to the known semi-Ohmic behavior. Lastly,
applying voltage between two bottom electrodes, Fig.~\ref{fig3}d, we obtain the fully
Ohmic behavior. This indicates that the bottom electrode is Ohmic. An equivalent
electrical scheme for our four-electrode configuration is presented in Fig.~\ref{fig3}a.
It is worthy of mentioning that removal of the original Cr electrode with subsequent
re-contacting leads to a pronounced non-Ohmic behavior. This observation excludes, in
particular, the possibility of leakage currents between the two bottom Cr electrodes.

The effect of heating on the CDW conductivity is an important issue. For temperature
measurements, we used a sample with thirty 1-mm-long domain lines and a silver-paste top electrode.
Up to $T = 70^{\,\circ}$C, the conductivity shows no signs of temporal
degradation within the period of about one day. For example, observation of $J(t)$ during $35$ hours at $T = 30^{\,\circ}$
indicates that the current fluctuates around an average value within only $1\%$, see
Fig.~\ref{fig4}a.
\begin{Figure}
\centering \vspace*{2mm}
\includegraphics[width=8.6cm]{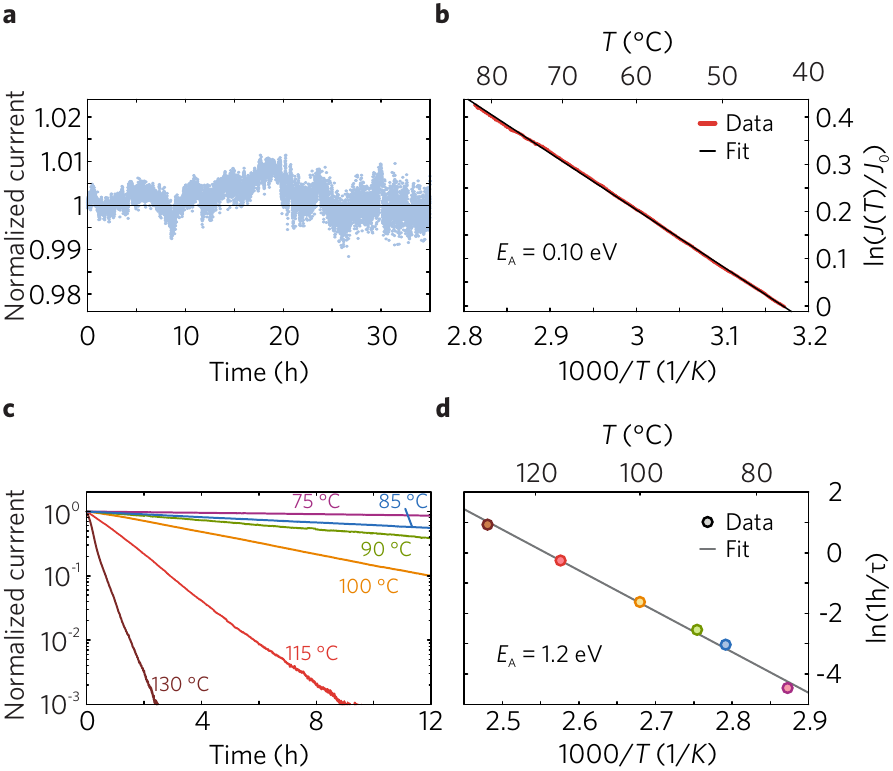}

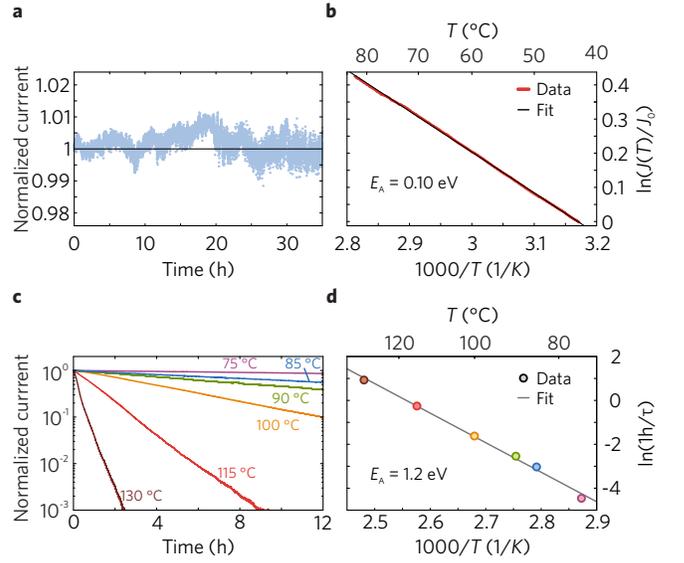
\captionof{figure}{{\bf Results of temperature-related measurements.}  {\bf a,} Temporal
dependence $J(t)$ at $30\,^{\circ}$C. {\bf b,} Arrhenius plot of steady-state value of
$J(T)$ for the temperature range $(40-85)\,^{\circ}$C. {\bf c,} Time decay of $J$ for $T
= 75,\,85,\,90,\,100, 115$, and $130\,^{\circ}$C. {\bf d,} Arrhenius plot of the decay
rate $\tau^{-1}$ for $T = (72-135)\,^{\circ}$C. }\label{fig4}
\end{Figure}
\noindent This contrasts with much stronger fluctuations of $J(t)$ in BaTiO$_3$ and
BiFeO$_3$. For $T \leq 70^{\,\circ}$, the current increases with temperature according to an Arrhenius law with
the activation energy $E_{\rm A} = (0.10 \pm 0.01)$~eV, see Fig.~\ref{fig4}b. This
temperature-activation dependence is the signature of hopping electron conductivity. For
temperatures $T > 70\,^{\circ}$C and on long time scales, the current decays exponentially in time,
$J(t)/J(0) = \exp(-t/\tau)$, see Fig.~\ref{fig4}c. The higher $T$, the faster is the
decay. Once the conductivity has vanished by heating, it cannot be recovered. The decay
rate $\tau^{-1}(T)$ also obeys an Arrhenius law with the activation energy $E_{\rm A} =
(1.2 \pm 0.1)$~eV, see Fig.~\ref{fig4}d. This value strongly indicates that migration of
some ions is responsible for the high-temperature decay of the CDW conductivity. 

Looking forward, it will be relevant to fully elucidate the effect reported here, in
particular with the aim to control all parameters: In-depth understanding is needed how
the inhomogeneous field employed for the calligraphic poling plus defects that pin the
domain walls influence the key outcome, the tilt of the domain walls. However, already
now it is clear that this effect can pave the way to novel integrated electronic-optical
devices: Calligraphically written charged domain walls can serve in a z-cut waver as
electrodes to access the electro-optical coefficient $r_{22}$, providing electrically
controlled local phase modulation. Patterning the wafer and contacting the electrodes
with nanometer-thin Ag wires allows the design of freely programmable optical phase
plates. Our findings will be also of use for integrated reconfigurable quantum-optical
devices providing, e.g., correlated photons from electro-optically tunable
difference-frequency generation. Flip-bonding a Si chip onto LN samples with tailored
conducting domain walls allows to apply freely designed 2D-structured electrical fields
to LN crystals, stimulating the field of LN-semiconductor hybrid chips.

 }

\end{multicols}
\vspace*{1mm}

{\footnotesize

\begin{multicols}{2}
\renewcommand\refname{

\end{multicols}

}


\vskip -1cm}
\begin{thebibliography}{1}

\bibitem{CatalanRMP12}
Catalan, G.,\ Seidel, J.,\ Ramesh,\ R., \& Scott, J.,\ F. Domain wall electronics, {\em
Reviews of Modern Physics} {\bf 84}, 119-156 (2012).

\vspace*{-1mm}

\bibitem{BednyakovSciRep15}
Bednyakov,\ P.\ S., Sluka, T.,\ Tagantsev,\ A.\ K.,\ Damjanovic,\ D.\ \& Setter, N.
Formation of charged ferroelectric domain walls with controlled periodicity. {\em Sci.
Rep.} {\bf 5}, 15819 (2015).

\vspace*{-1mm}

\bibitem{FeiglSciRep16}
Feigl, L.,\ Sluka,\ T.,\ McGilly, L.\ J.,\ Crassous, A.,\ Sandu,\ C.\ S.\ \& Setter,\ N.
Controlled creation and displacement of charged domain walls in ferroelectric thin
films. {\em Sci.\ Rep.} {\bf 6}, 31323 (2016).

\vspace*{-1mm}

\bibitem{SeidelPCL12}
Seidel,\ J. Domain walls as nanoscale functional elements. {\em J.\ Phys.\ Chem.\ Lett.}
{\bf 3}, 2905-2909 (2012).

\bibitem{VasudevanAFM13}
Vasudevan, R.\ K. {\em et al.}\ Domain wall conduction and polarization-mediated
transport in ferroelectrics. {\em Adv.\ Funct.\ Mater.} {\bf 23}, 2592-2616 (2013).

\vspace*{-1mm}

\bibitem{Vul73}
Vul, B.\ M.,\ Guro, G.\ M.\ \& Ivanchik,\ I.\ I. Encountering domains in ferroelectrics.
{\em Ferroelectrics} {\bf 6}, 29-31 (1973).

\vspace*{-1mm}

\bibitem{Maksymovych12}
Maksymovych, P.\ {\em et al.} Tunable metallic conductance in ferroelectric nanodomains.
{\em Nano Lett.} {\bf 12}, 209-2013 (2012).

\vspace*{-1mm}

\bibitem{VasudevanNL12}
Vasudevan, R.\ K.\ {\em et al}, Domain wall geometry controls conduction in
ferroelectrics. {\em Nano.\ Lett.} {\bf 12}, 5524-5531 (2012).

\vspace*{-1mm}

\bibitem{SlukaNC13}
Sluka, T., Tagantsev, A.\ K., Bednyakov, P.\ \& Setter, N. Free-electron gas at charge
domain walls in insulating BaTiO$_3$. {\em Nat.\ Commun.} {\bf 4}, 1808 (2013).

\vspace*{-1mm}

\bibitem{Chapter16}
Sluka, T., Bednyakov,\ P., Yudin,\ P., Crassous, A.,\ \& Tagantsev, A. Charged domain
walls in ferroelectrics, chapter in Topological structures in ferroic materials - domain
walls, vortices, and skyrmions. Springer Series in Material Science {\bf 228}, 103-138
(2016).

\vspace*{-1mm}

\bibitem{SoergelAFM12}
Schr\"{o}der, M.\ {\em at al.} Conducting domain walls in lithium niobate single
crystals. {\em Adv.\ Funct.\ Mater.} {\bf 22}, 3936-3944 (2012).

\vspace*{-1mm}

\bibitem{Arizmendi04}
Arizmendi,\ L. Photonic applications of lithium niobate crystals. {\em Phys.\ Stat.\
Sol.} {\bf 201}, 253–283 (2004).

\bibitem{ShurAPL13}
Shur, V.\ Ya., Baturin,\ I.\ S., Akhmatkhanov, A.\ R., Chezganov,\ D.\ S.\ \& Esin, A.\
A. Time-dependent conduction current in lithium niobate crystals with charged domain
walls. {\em Appl.\ Phys.\ Lett.} {\bf 103}, 102905 (2013).

\vspace*{-1mm}

\bibitem{Dunn99}
Dunn,\ M.\ H.,\ \& Ebrahimzadeh,\ M. Parametric generation of tunable light from
continuous-wave to femtosecond pulses. {\em Science} {\bf 286}, 1513-1517 (1999).

\vspace*{-1mm}

\bibitem{WeigelSciRep16}
Weigel,\ P.\ O.\ {\em et al.} Lightwave circuits in lithium niobate through hybrid
waveguides with silicon photonics. {\em Sci. Rep.} {\bf 6}, 22301 (2016).

\vspace*{-1mm}

\bibitem{GurievPRB11}
Guriev,\ M.\ Y.,\ Tagantsev,\ A.\ K.\ \& Setter,\ N. Head-to-head and tail-to-tail
180$^{\circ}$ domain walls in an isolated ferroelectric. {\em Phys.\ Rev. B} {\bf 83},
184101 (2011).

\vspace*{-1mm}

\bibitem{Zuo14}
Zuo,\ Y.\ Genenko,\ Y.\ A. \& Xu,\ B.-X. Charge compensation of head-to-head and
tail-to-tail domain walls in barium titanate and its influence on conductivity. {\em J.
Appl. Phys.} {\bf 116}, 044109 (2014).

\vspace*{-1mm}

\bibitem{ShurPRB11}
Eliseev,\ E.\ A.,\ Morozovska,\ A.\ N.,\ Svechnikov,\ G.\ S.,\ Gopalan,\ V., \& Shur,\
V.\ Ya. Staic conductivity of charged domain walls in uniaxial ferroelectric
semiconductors. {\em Phys.\ Rev.\ B} {\bf 83}, 235313 (2011).

\vspace*{-1mm}

\bibitem{Grekov76}
Grekov,\ A.\ A., Adonin,\ A.\ A., \& Protsenko,\ N.\ P. Encountering domains in SbSI.
{\em Ferroelectrics} {\bf 13}, 483-485 (1976).

\vspace*{-1mm}

\bibitem{SeidelNM09}
Seidel,\ J.\ {\em et al.} Conduction at domain walls in oxide multiferroics. {\em Nat.\
Mat.} {\bf 8}, 229-234 (2009).

\vspace*{-1mm}

\bibitem{Crassous15}
Crassous,\ A., Sluka,\ T., Tagantsev,\ A., \& Setter,\ N. Polarization charge as a
reconfigurable quasi-dopant in ferroelectric thin films, {\em Nat.\ Nanotechnol.} {\bf
10}, 614-618 (2015).

\vspace*{-1mm}

\bibitem{Bednyakov16}
Bednyakov,\ P.\ Sluka,\ T.,\ Tagantsev,\ A.,\, Damjanovic,\ D. \& Setter,\ N.
Free-carrier-compensated charged domain walls produced with super-bandgap illumination
in insulating ferroelectrics. {\em Adv.\ Mater.} DOI: 10.1002/adma.201602874 (2016).

\vspace*{-1mm}

\bibitem{MeierNM12}
Meier,\ D.\ {\em et al.} Anisotropic conductance at improper ferroelectricdomain walls.
{\em Nat\ Mat.} {\bf 11}, 284-288 (2012).

\vspace*{-1mm}

\bibitem{Wu12}
Wu,\ W.,\ Horibe,\ Y.,\ Lee,\ N., Cheong,\ S.-W.\ \& Guest,\ J.\ R. Conduction of
topologically protected charged ferroelectric domain walls. {\em Phys.\ Rev.\ Lett.}
{\bf 108}, 077203 (2012).

\vspace*{-1mm}

\bibitem{ShurAPL13}
Shur,\ V.\ Ya.,\ Baturin,\ I.\ S.,\ Akhmatkhanov,\ A.\ R.,\ Chezganov,\ D.\ S. \& Esin,
A.\ A. Time-dependent conduction current in lithium niobate crystals with charged domain
walls. {\em Appl.\ Phys.\ Lett.} {\bf 103}, 102905 (2013).

\vspace*{-1mm}

\bibitem{LNPoling}
Terabe,\ K. {\em et al.} Microscale to nanoscale ferroelectric domain and surface
engineering of a near-stoichiometric LiNbO$_3$ crystal. {\em Appl.\ Phys.\ Lett.} {\bf
82}, 433–435 (2003).

\bibitem{CallPoling05}
Mohageg,\ M.\ {\em et al.} Calligraphic poling of lithium niobate. {\em Opt.\ Express}
{\bf 13}, 3408-3419 (2005).

\vspace*{-1mm}

\bibitem{LNProperties}
Properties of Lithium Niobate, K.\ K.\ Wong, ed.\ (2002).

\bibitem{LiNL13}
Li,\ L.\ {\em et al.} Atomic scale structure changes induced by charged domain walls in
ferroelectric materials. {\em Nano.\ Lett.} {\bf 13}, 5218-5223 (2013).

\vspace*{-1mm}

\bibitem{Smythe}
Smythe,\ W.\ R. {\em Static and Dynamic Electricity.} (McGraw-Hill, 1950).


\end{thebibliography}
\end{document}